\documentclass[runningheads]{llncs}
\usepackage{amsmath}
\usepackage{adjustbox}
\usepackage{hyperref,cleveref}
\usepackage{booktabs,rotating}
\usepackage{comment}
\usepackage{graphicx}
\usepackage{multirow}
\usepackage{placeins}
\usepackage{float}
\usepackage{caption}
\usepackage{subcaption}
\usepackage{siunitx}
\usepackage{layouts}

\usepackage{footnote}
\usepackage{tablefootnote}
\usepackage{threeparttable}

\usepackage{tabularx,booktabs}
\newcolumntype{Y}{>{\centering\arraybackslash}X}

\usepackage{hyperref,cleveref}
\hypersetup{colorlinks=true,linkcolor=blue,urlcolor=blue,citecolor=blue}

\usepackage{xcolor}

\begin{document}

\title{Mandibular Teeth Movement Variations in Tipping Scenario: A Finite Element Study on Several Patients}

\titlerunning{Mandibular Teeth Movement Variations in Tipping Scenario}

\authorrunning{T.Gholamalizadeh et al.}

\author{Torkan Gholamalizadeh \inst{1,2} \and 
        Sune Darkner \inst{2} \and 
        Paolo Maria Cattaneo\inst{3} \and 
        Peter Søndergaard \inst{1} 
        \and Kenny Erleben \inst{2}}
        
\institute{3Shape A/S, Copenhagen, DK, \\\and
Department of Computer Science, University of Copenhagen, Copenhagen, DK, \\\and
Department of Dentistry and Oral Health, Aarhus University, Aarhus, DK 
\email{\{torkan, darkner, kenny\}@di.ku.dk}\\  \email{paolo.cattaneo@dent.au.dk }\\
\email{peter.soendergaard@3shape.com}}

\maketitle 

\begin{abstract}

Previous studies on computational modeling of tooth movement in orthodontic treatments are limited to a single model and fail in generalizing the simulation results to other patients. To this end, we consider multiple patients and focus on tooth movement variations under the identical load and boundary conditions both for intra- and inter-patient analyses. We introduce a novel computational analysis tool based on finite element models (FEMs) addressing how to assess initial tooth displacement in the mandibular dentition across different patients for uncontrolled tipping scenarios with different load magnitudes applied to the mandibular dentition. This is done by modeling the movement of each patient's tooth as a nonlinear function of both load and tooth size. As the size of tooth can affect the resulting tooth displacement, a combination of two clinical biomarkers obtained from the tooth anatomy, i.e., crown height and root volume, is considered to make the proposed model generalizable to different patients and teeth.


\end{abstract}

\section{Introduction}

\begin{figure*}[t]
  \centering
  \includegraphics[width=0.92\textwidth]{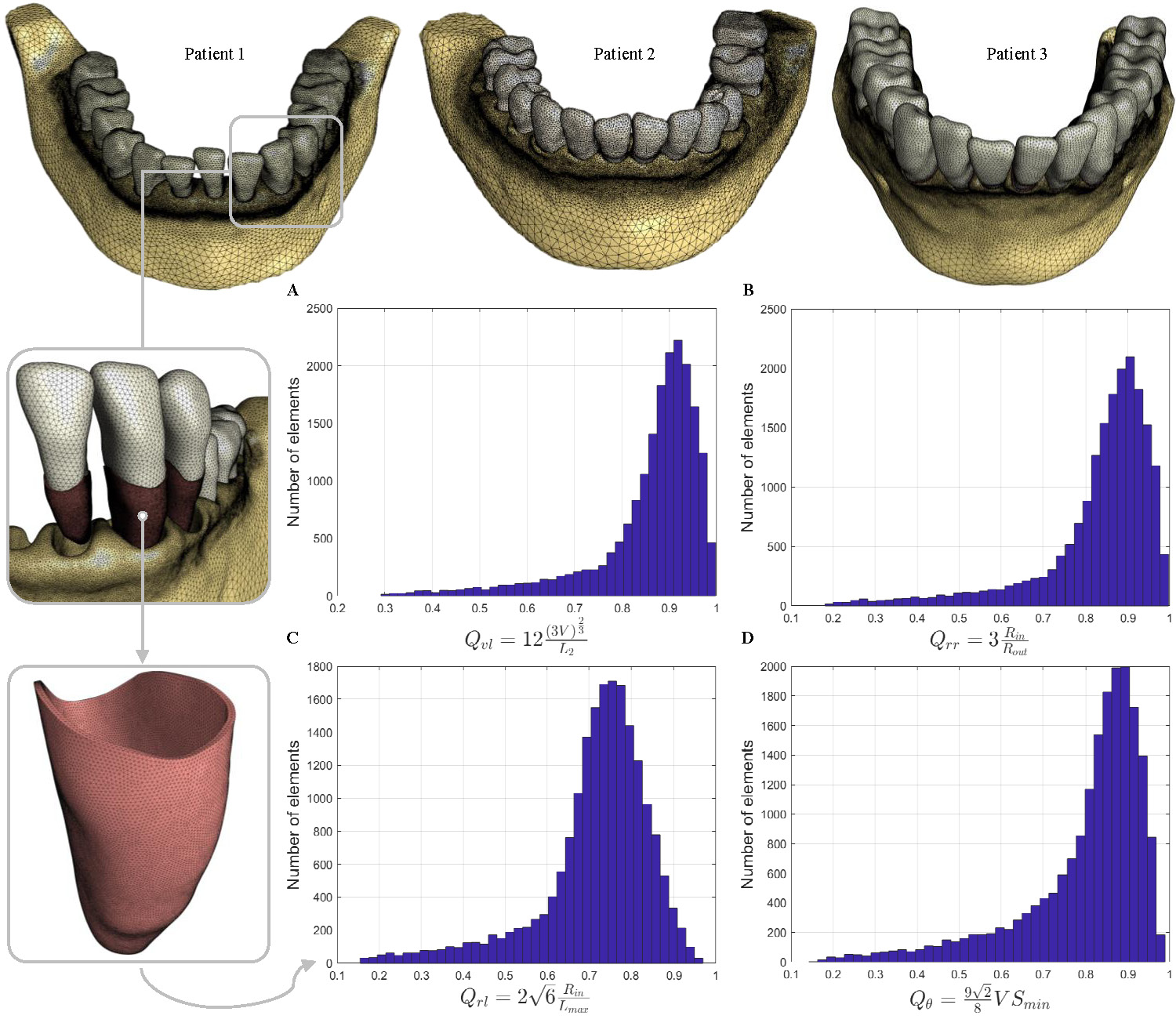}
  \vspace{-0.15cm}
  \caption{FEMs and meshes for three patients, and mesh quality histograms for the PDL of the left canine. \textbf{A:} The volume-edge ratio, \textbf{B:} The radius ratio, \textbf{C:} The radius-edge ratio, \textbf{D:} A metric introduced in \cite{freitag1997tetrahedral}.}
  \label{fig:population_mesh}
  \vspace{-0.3cm}
\end{figure*}

Orthodontic tooth movement is the result of alveolar bone remodeling caused by the applied forces and deformations in the periodontium. Finite element models (FEMs) is widely used to assess stress/strain in the alveolar bone and periodontal ligament (PDL), the fibrous connective tissue between tooth and bone, in the orthodontic treatments \cite{cattaneo2005finite,chen2014periodontal,savignano2016nonlinear,likitmongkolsakul2018development,hamanaka2017numeric,kawamura2019biomechanical}. Moreover, the initial and long-term tooth movements can be investigated using these models. In this work, we use an FEM to provide a biomechanical model of the full mandibular dentition focusing on initial teeth displacements caused by the applied load on the teeth (see \Cref{fig:population_mesh} and \Cref{fig:Tooth_PDE}). We generate patient-specific FEMs for three patients by segmenting the cone-beam computed tomography (CBCT) scans of the patients. The models are provided by using the same boundary conditions under the same scenarios. In each scenario, an identical force magnitude is applied perpendicular to the surface of each tooth to mimic an uncontrolled tipping movement. Besides, the load magnitude can change from \SIrange{0.3}{1}{\newton}  with \SIlist{0.1}{\newton} increments, and teeth transformations are recorded for all teeth of each patient. Finally, the results are compared with the corresponding teeth of other patients.

Our hypothesis is that variations in the teeth anatomy of different patients and the load magnitudes can affect the resulting tooth displacement. Therefore, in this study, tooth movements (i.e., rotation and translation) are estimated as nonlinear functions of both load and the ratio of crown height to root volume using the obtained biomechanical models.

\section{Related Work}

The performance of the orthodontic treatments can be improved if the movement of the teeth could be predicted in a reliable way. Therefore, many studies have focused on predicting tooth movements in orthodontic treatments using FEMs. In general, the tooth movement occurs in two phases \cite{hamanaka2017numeric}. In the first phase \cite{liang2009torque,cattaneo2008moment,cattaneo2005finite,savignano2016nonlinear}, which is the main focus of this study, tooth moves within the PDL space in few seconds after applying a force \cite{li2018orthodontic}. This movement is substantially due to the deformation of the PDL tissue caused by the applied load. In the second phase \cite{likitmongkolsakul2018development,hamanaka2017numeric,kawamura2019biomechanical,likitmongkolsakul2018development,chen2014periodontal}, the resulting stress in the PDL and bone tissue causes a bone remodeling process, where the bone is resorbed and formed in the compressed and stretched regions of the PDL, respectively.

In the context of FE-based modeling of the initial tooth movement, some studies \cite{cattaneo2008moment,savignano2016nonlinear} have investigated different types of movements individually including bodily movement, controlled tipping, and uncontrolled tipping. Some others have explored the teeth mesialization, distalization, or retraction scenarios \cite{liang2009torque,kawamura2019biomechanical,likitmongkolsakul2018development,park2017biomechanical}. These studies have considered the effect of the force direction \cite{savignano2016nonlinear,kawamura2019biomechanical},
moment-to-force \cite{cattaneo2008moment,savignano2016nonlinear}, and force magnitude \cite{melsen2007importance,cattaneo2008moment} on tooth transformation \cite{cattaneo2008moment,kawamura2019biomechanical,park2017biomechanical} or location of the center of rotation \cite{cattaneo2008moment,savignano2016nonlinear}. However, the jaw model, force system, and number of teeth used in the analyses are not consistent. For example, \cite{cattaneo2008moment,savignano2016nonlinear} used a small portion of jaw, while \cite{huang2020mandible} worked on a fully segmented jaw model. Likewise, different studies have examined different number of teeth, e.g., using a single tooth \cite{savignano2016nonlinear}, two \cite{cattaneo2008moment,liang2009torque} or more \cite{kawamura2019biomechanical,park2017biomechanical}. The force and/or moments have also been applied to different parts including the surface of tooth \cite{cattaneo2008moment,liang2009torque}, center of the resistance \cite{savignano2016nonlinear}, and orthodontic appliances \cite{kawamura2019biomechanical,likitmongkolsakul2018development}. 

The abovementioned biomechanical models, however, might not be applicable for analyzing different teeth motions obtained from multiple patients. In other words, the obtained tooth displacement results represented only by visualizing the displacement fields \cite{liang2009torque,park2017biomechanical}, measuring the displacement of the selected landmarks \cite{cattaneo2008moment,park2017biomechanical}, or acquiring the translations/rotations using some predefined measurement points \cite{kawamura2019biomechanical} lack useful information about different tooth motion tendencies for full dentition of multiple patients and, hence, are less interpretable when it comes to the across patients modeling analyses.

Moreover, existing FEMs applied in computational orthodontics are mostly limited to a single patient \cite{cattaneo2008moment,liang2009torque,savignano2016nonlinear,kawamura2019biomechanical,huang2020mandible,park2017biomechanical}. Although Likitmongkolsakul et al. \cite{likitmongkolsakul2018development} propose a stress-movement function of a canine for two orthodontic patients under an identical scenario, to the best of our knowledge, there are no other studies considering multiple patients for tooth movement modeling.

In this work, by considering the biomechanical models of human mandible acquired from CBCT scans of three patients, we investigate the tooth movement variations in multiple patients using rigid body transformations under different load magnitudes. To the best of our knowledge, this is the first computational model in orthodontics applied to three different patients. Our experiments consist of both intra- and inter-patient analyses. Considering teeth motions under an identical scenario, both in intra-patient and inter-patient analyses, helps us to obtain a general pattern for the movement of different teeth using patient-specific teeth and bone geometries.

\section{Setting up the Finite Element Model}

This section describes different consecutive steps that are conducted to generate a patient-specific FEM of the human mandible. First, the geometry reconstruction takes place by segmenting the CBCT scan of the patient. Second, the surface mesh of the obtained geometries are re-meshed and a volumetric mesh is generated for each geometry. Next, the resulting volumetric meshes are imported into a finite element (FE) framework to set up the FE problem. The details of the biomechanical model, e.g., boundary conditions, contact definitions, and utilized material models are presented in this section. Finally, the model is numerically verified by using mesh convergence study and parameter sensitivity analysis.

Segmentation is performed using 3D Slicer \cite{fedorov20123d} based on a semi-automatic watershed algorithm applied to the bone and teeth. Next, the wrongly segmented regions are modified to obtain the final segmentation result. Since the resolution of the orthodontic scans with a voxel size of \SI[product-units=single]{0.3 x 0.3 x 0.3}{mm^3} is not high enough for segmenting the thin PDL tissue ($\approx$ \SIlist{0.2}{\milli\metre} width) from the scans, the PDL layer is generated with a uniform width of \SIlist{0.2}{\milli\metre} around each tooth root as shown in \Cref{fig:population_mesh}. We select three patients' scans of various crown height, root length, and teeth sizes, to ensure having enough geometrical variations. Each segmentation result is later verified by an orthodontic expert.

The segmented geometries are exported as surface meshes in STL files. These meshes are decimated and re-meshed using Meshmixer \cite{Meshmixer} to provide high-quality surface meshes. Uniform meshes are used for teeth and PDL geometries. \Cref{table:model_params} presents the edge length of the triangular meshes for each component. For bone geometry, an adaptive mesh is generated in which the edge length of the surface mesh triangles varies between \SIlist{0.4}{\milli\metre} and \SIlist{2}{\milli\metre} from the neighboring regions to the PDL and the bottom region of the mandible. Utilizing an adaptive mesh helps us to obtain a finer mesh in the regions of interest, and consequently, an accurate result in the FE analysis, yet reducing the total number of elements.

\begin{table}[!b]
\centering
\renewcommand{\arraystretch}{1.5}
\caption{Summary of the materials and mesh properties.}
\label{table:model_params}
\resizebox{\textwidth}{!}{%
\begin{tabular}{l|l|cc|lclrl}
\hline
 & \multirow{2}{*}{Material model} & \multicolumn{2}{c|}{Material properties} &  & Mesh Properties &  & \multicolumn{1}{l}{} & \textbf{} \\
 \cline{3-4} \cline{5-9} 
 &  & \begin{tabular}[c]{@{}c@{}}Young's Modulus\\ 
 (MPa)\end{tabular} & \begin{tabular}[c]{@{}c@{}}Poisson's Ratio\\ (-)\end{tabular} &  & \begin{tabular}[c]{@{}c@{}}Surface mesh \\ edge length (mm)\end{tabular} &  & \multicolumn{1}{l}{\begin{tabular}[c]{@{}l@{}}Number of\\  tetrahedra\end{tabular}} & \textbf{} \\ \hline
Tooth & Rigid Body & - & - &  & 0.4  &  & 8,000 & \multirow{4}{*}{} \\
Bone & Isotropic elastic & $1.5\times10^{3}$ & 0.3 &  & \begin{tabular}[c]{@{}c@{}}Adaptive mesh\\ (from 0.4 to 2)\end{tabular} &  & 3,255,000 &  \\ \cline{3-4}
 &  & $C_1$ (\SIlist{}{\mega\pascal}) & $C_2$ (\SIlist{}{\mega\pascal}) &  &  & \multicolumn{1}{l}{} &  &  \\ \cline{3-4}
PDL & Mooney-Rivlin $^*$ & 0.011875 & 0 &  & 0.1 &  & 90,700 &  \\ \hline
\end{tabular}%
}
\begin{tablenotes}
\item{\footnotesize $^{*}$ $C_2 = 0$ reduces the Mooney-Rivlin material to uncoupled Neo-hookean. The values assigned for $C_1$ and $C_2$ correspond to the Young’s modulus and Poisson's ratio of \SIlist{0.0689}{\mega\pascal} and $0.45$, respectively.}
\end{tablenotes}
\vspace{-0.2cm}
\end{table}

\begin{figure}[t]
  \centering
  \includegraphics[width=0.98\linewidth]{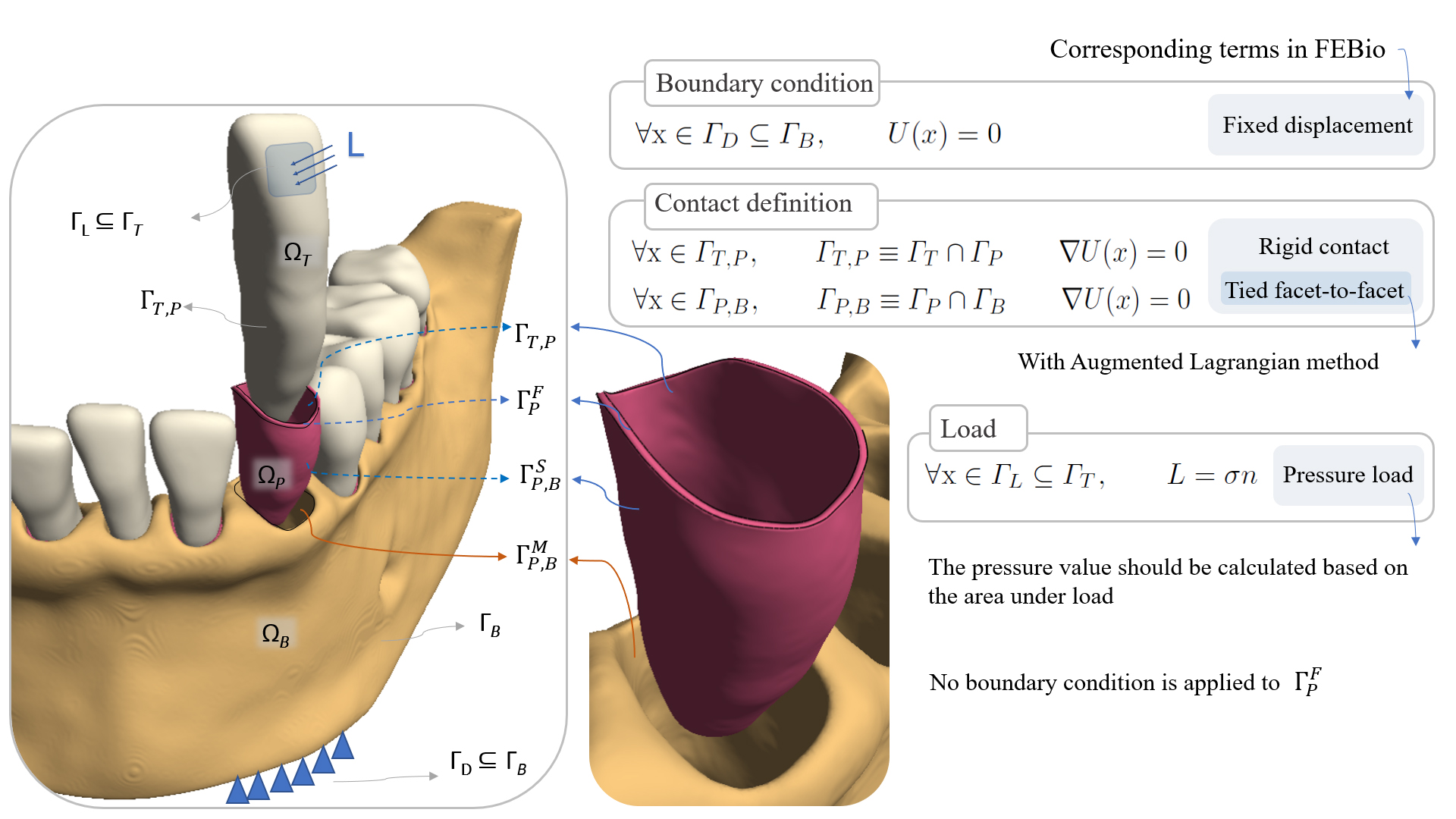}
  \vspace{-0.1cm}
  \caption{A closeup view of 
  the tooth supporting complex. The contacts between different domains and boundary conditions are presented. Tooth ($\Omega_{T}$), Periodontal ligament ($\Omega_{P}$), Alveolar bone ($\Omega_{B}$), Tooth-PDL contact ($\Gamma_{T,P}$), PDL-Bone contact ($\Gamma_{P,B}$) and a Dirichlet boundary condition 
  ($\Gamma_{D}$) are shown.}
  \label{fig:Tooth_PDE}
  \vspace{-0.4cm}
\end{figure}

High quality volumetric meshes are generated for each surface mesh using TetGen \cite{TetGen}, by defining an upper limit for the radius-edge ratio of to-be-generated tetrahedra. This mesh quality constraint controls the ratio between the radius of the circumscribed sphere and the shortest edge of each tetrahedron, which prevents the production of low-quality (badly shaped) tetrahedra. Later, four different mesh quality measurements presented in \cite{shewchuk2002good}, i.e., the volume-edge ratio \cite{misztal2013multiphase,liu1994relationship}, the radius-edge ratio \cite{baker1989element}, and the radius ratio \cite{caendish1985apporach,freitag1997tetrahedral} are chosen to verify the quality of the generated 4-noded tetrahedral meshes (TET4) (see the quality histograms in \Cref{fig:population_mesh}). Finally, the obtained meshes are used to set up and solve the FE problem. For reproducibility, we generate and solve the computational biomechanical models in FEBio software package \cite{FEBio} which is an open-source software for nonlinear FEA in biomechanics. A nonlinear quasi-static simulation is performed to analyze the teeth displacements in each FE model.

The different domains of the FEM, material properties, contact types, boundary conditions and the applied load are summarized in \Cref{fig:Tooth_PDE}. To simplify the proposed model, the tooth domain is assumed as rigid-body with 6 degrees of freedom. The center of mass for each tooth is calculated automatically using FEBio based on a predefined density parameter \cite{FEBioUserManual}. Furthermore, since the deformation of the bone tissue is negligible under the orthodontic forces, no distinction is made for the cortical and trabecular bone \cite{ziegler2005numerical,qian2009deformation}, and an isotropic elastic material model is used for the homogeneous bone geometry.

The importance of the PDL tissue in transferring loads from the tooth to the alveolar bone has been shown in the literature \cite{mccormack2014biomechanical,ortun2018approach,cattaneo2005finite}. Accordingly, the PDL tissue is included in our model as a thin layer of finite elements \cite{ortun2018approach,hohmann2011influence,cattaneo2005finite,savignano2016nonlinear}. This allows for investigating the stress/strain field in the PDL, e.g., using data-driven models, that can later be used in the bone remodeling process \cite{chen2014periodontal}. Moreover, several studies have characterized the biomechanical behavior of the PDL tissue \cite{dorow2003determination,uhlir2016biomechanical,qian2009deformation}, some of which have suggested the Mooney-Rivlin Hyperelastic (MRH) model for the PDL \cite{qian2009deformation,uhlir2016biomechanical}. In this study, an MRH material model is used for the PDL domain based on the parameter values reported in \Cref{table:model_params}.

The Tooth-PDL interface and PDL-Bone interface are fixed in both normal and tangent directions using a Neuman condition (see \Cref{fig:Tooth_PDE}). In addition, all elements at the bottom surface of the bone ($\Gamma_{D}$) are fixed in all directions by applying a Dirichlet boundary condition.

To mimic the uncontrolled tipping scenario, a pressure load is applied perpendicular to the labial/buccal surface of the tooth crown, as shown in \Cref{fig:Tooth_PDE}. The area under the load, which represents the area under the orthodontic bracket, is set to the center of the teeth crowns. To ensure that the same force magnitude is applied to the teeth, the area under the load is measured separately for each tooth. Next, the corresponding pressure value for the desired force magnitude is calculated and used as the \textit{pressure load} in FEBio. An identical force magnitude is exerted to all teeth simultaneously in order to investigate the tooth movement variations of the mandibular teeth across the three patients.

\begin{figure}[t]
  \centering
  \includegraphics[width= 0.8\linewidth]{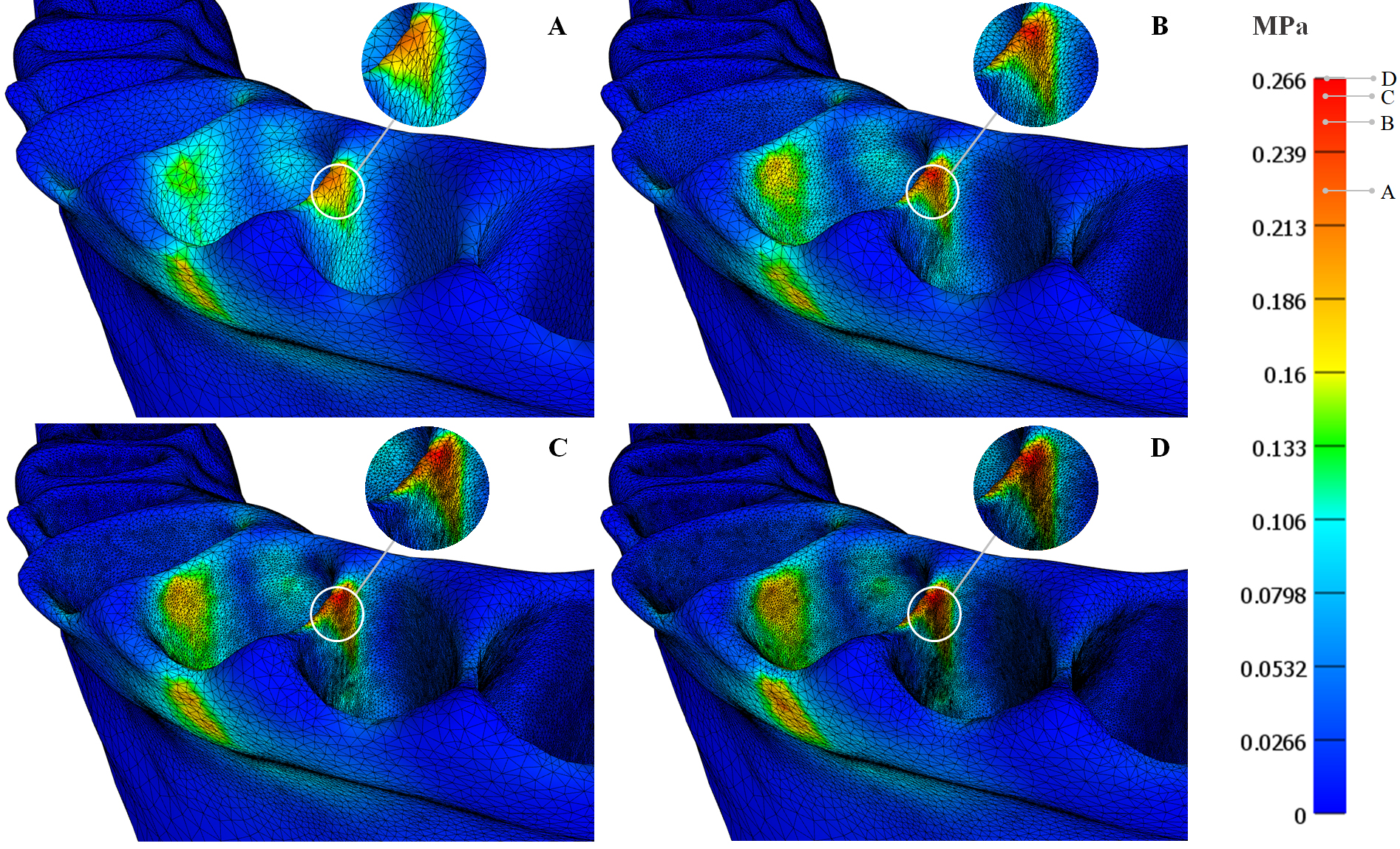}
  \vspace{-0.1cm}
  \caption{Mesh convergence study showing the Von Mises stress in the bone geometry under the same boundary conditions. \textbf{A to D:} The model with $N$ (coarse), $2N$, $4N$, and $8N$ elements. The stress fields are consistent in the finer models.}
  \label{fig:Conv_study_models}
\end{figure}

The model is then verified by studying the mesh convergence and parameters sensitivities. The final resolution of the mesh is defined in the mesh convergence study process where the total number of elements, except for the rigid body teeth, is iteratively increased by a factor of 2 until the relative error is less than 4\% of the maximum stress (see \Cref{fig:Conv_study_models}). The number of tetrahedra in the refined mesh is presented in \Cref{table:model_params}. The parameter sensitivity study, summarized in \Cref{table:sensitivity}, is done on the material parameters of different tissues and the parameters used for the tied contact in the PDL-bone interface.

\begin{table}[!t]
\renewcommand{\arraystretch}{1.4}
\centering
\caption{Summary of parameter sensitivity analysis conducted on the model.}
\label{table:sensitivity}
\resizebox{\textwidth}{!}{%
\begin{tabular}{c|c|c|c|c|c|c}
\hline
\multirow{2}{*}{} & \multicolumn{2}{c|}{PDL} & \multicolumn{2}{c|}{Bone} & \multicolumn{2}{c}{PDL-Bone Tied Contact} \\ \cline{2-7} 
 & Young's Modulus & Poisson's Ratio & Young's Modulus & Poisson's Ratio & \begin{tabular}[c]{@{}c@{}}Augmented\\  Lagrangian\end{tabular} & Penalty Factor \\ \hline
\multicolumn{1}{l|}{\begin{tabular}[c]{@{}c@{}}Interval of \\ parameter change\end{tabular}} & 0.044 - 0.0938 & 0.45 - 0.49 & 1200 - 13700 & 0.2 - 0.4 & 0.2 - 0.1 & 0.25 - 1.75 \\ \hline
\begin{tabular}[c]{@{}c@{}}Relative difference in\\ Von Mises stress (\%)\end{tabular} & 2.127 & 2.127 & 0.709 & 3.900 & 1.418 & 3.900 \\ \hline
\end{tabular}%
}
\end{table}

\section{Experiments and Results}

This section describes the experimental setup under the uncontrolled tipping scenario among the three patients. First, the obtained results are presented and discussed. Next, we propose two functions that describe the translation and rotation of different tooth types of the patients based on tooth IDs and selected clinical biomarkers, i.e., crown height and root volume.

\begin{figure}[!t]
\vspace{-0.4cm}
	\centering
	\includegraphics[width=0.45\linewidth]{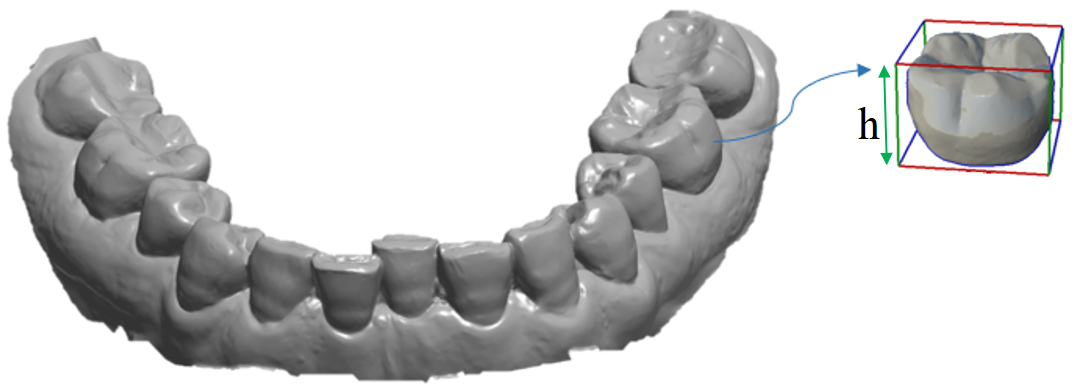}
 	\vspace{-0.2cm}
 	\caption{An intraoral scan of a patient and the obtained crown height for a tooth.}
 	\label{fig:trios}
 	\vspace{-0.3cm}
\end{figure}

In order to have a comprehensive inter-patient analysis, we select the patients with roughly the same number of teeth. We use the intraoral scan of each patient, captured by the 3Shape Trios scanner \cite{3ShapeTrios}, to obtain the crown height of each tooth. Therefore, we ensure that the intraoral optical scans are available for all patients, and the CBCT scans have sufficient quality for performing the segmentation. \Cref{fig:trios} shows an intraoral scan of a patient and illustrates how the crown height is obtained for a tooth.

\begin{figure}[t]
	\centering
	\includegraphics[width=0.99\linewidth]{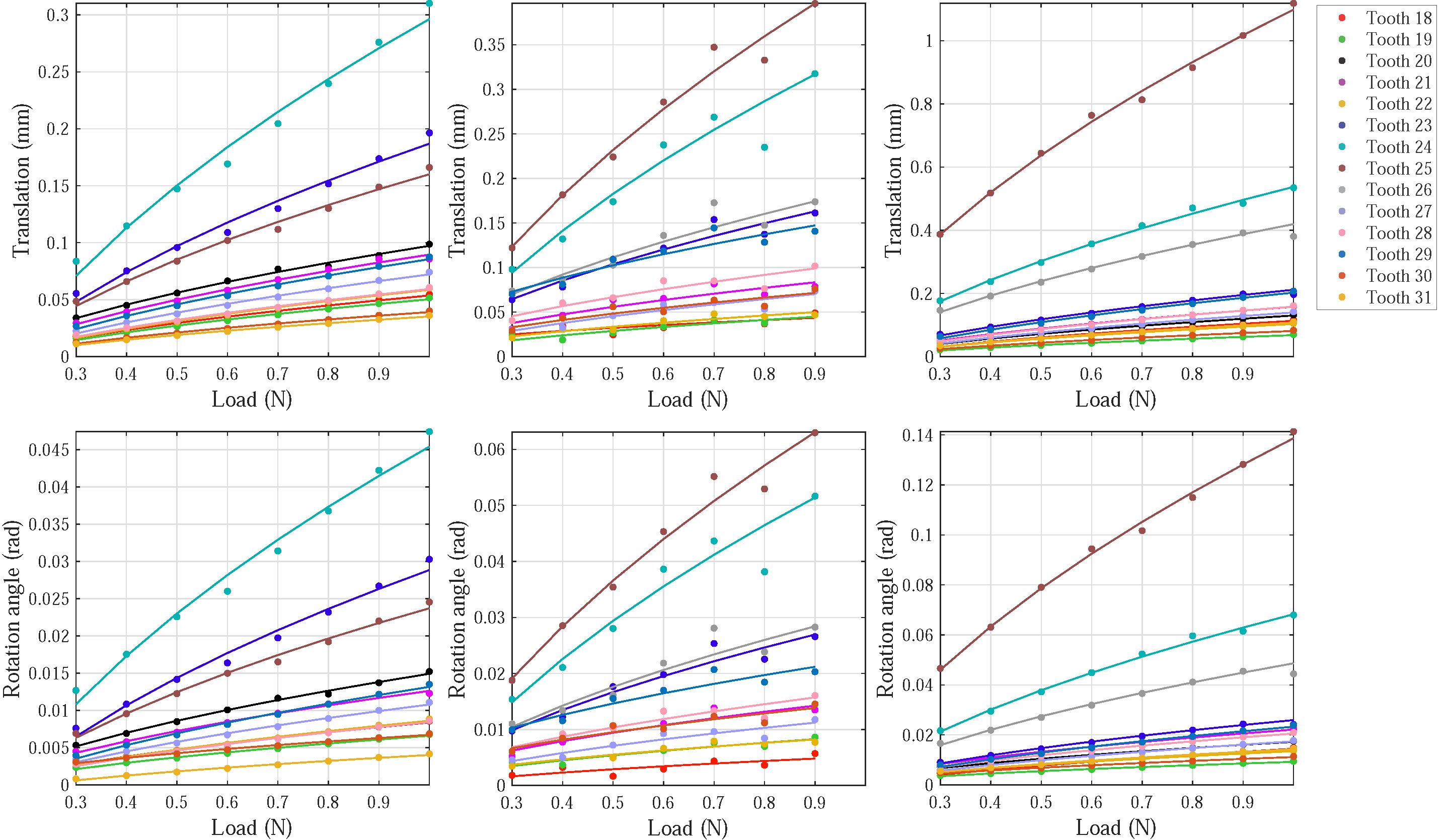}
	\vspace{-0.15cm}
 	\caption{The translation magnitudes and rotation angles versus the applied load. A nonlinear square root regression model is applied to fit the data from each patient's tooth. \textbf{Top-row:} Translation magnitude, \textbf{Bottom-row:} Rotation angle, \textbf{Left to Right:} Patient 1 through 3.}
 	\label{fig:result_regs}
 	\vspace{-0.1cm}
\end{figure}

In each scenario, an identical force magnitude is applied perpendicular to the surface of each tooth. The load magnitude ($l$) varies from \SIrange{0.3}{1}{\newton} with \SIlist{0.1}{\newton} intervals. The displacement of each tooth is measured as the translation of  center of the mass ($\vec t$) and rotation of the rigid body teeth (with angle $\theta$ and axis $\vec n$). Besides, in each simulation, we record the tooth ID ($k$), load magnitude, and the relevant biomarkers. \textit{Universal Numbering (UNN)} system is used for the tooth ID, where $k$ changes between $17$ and $32$ from the left third molar to the right third molar, respectively. \Cref{fig:result_regs} illustrates relation between the translation/rotation and applied load for different teeth of each patient. As can be seen, the translation magnitude of the mandibular incisors for an applied load of \SIlist{0.4}{\newton} changes from \SIrange{0.07}{0.11}{\milli\metre}, \SIrange{0.13}{0.18}{\milli\metre}, and \SIrange{0.24}{0.51}{\milli\metre}, for patient 1 through 3, respectively. These values are similar to the results of the clinical study done by Jones et al. \cite{jones2001validated}, where the obtained initial tooth movements ranged from \SIrange{0.012}{0.133}{\milli\metre} for maxillary incisors of ten patients under a constant load of \SIlist{0.39}{\newton} over one-minute cycles.

The translation magnitude and rotation angle of the teeth can be described as the square root functions of the applied load, i.e., $t_{j,k} = \alpha_{t_{j,k}} \sqrt{l \,} + \beta_{t_{j,k}}$ and $\theta_{j,k} = \alpha_{\theta_{j,k}} \sqrt{l \,} + \beta_{\theta_{j,k}}$, where $\alpha_{t_{j,k}}$ and $\alpha_{\theta_{j,k}}$ are the translation/rotation function coefficients for the $k$-th tooth of the $j$-th patient, and $\beta_{t_{j,k}}$ and $\beta_{\theta_{j,k}}$ are the corresponding function intercepts which are nearly zero. This nonlinear relation between the displacement and  load is in line with the experimental findings of the clinical study of \cite{christiansen1969centers} and numerical results of the biomechanical model of \cite{cattaneo2008moment}. However, the function coefficients vary across different patients' teeth, i.e., the values increase when moving from the molars to central incisors.

It can be deduced from \Cref{fig:result_regs} that the estimated coefficients of the nonlinear functions vary across different patients for the same tooth types. This finding is due to the fact that the initial tooth movement can be influenced by different factors such as tooth anatomical variations and surrounding alveolar bone and PDL layer. Additionally, the root length of tooth and its surrounding alveolar bone can affect the initial tooth movement, center of rotation, and center of the resistance \cite{tanne1991patterns}. The same behavior applies for the crown height. In other words, a specific tooth with a longer crown (or a shorter root) would experience more displacements than the same tooth with a shorter crown (or a longer root). However, the exact relationship between the crown/root size and tooth displacement is missing. Our hypothesis is that the intra- and inter-patient variations in crown and root size can influence the teeth movements of different patients. Therefore, we propose the ratio of crown height to root volume as the biomarker causing tooth movement variations together with the applied load.

\begin{figure}[t]
	\centering
	\includegraphics[width=0.95\linewidth]{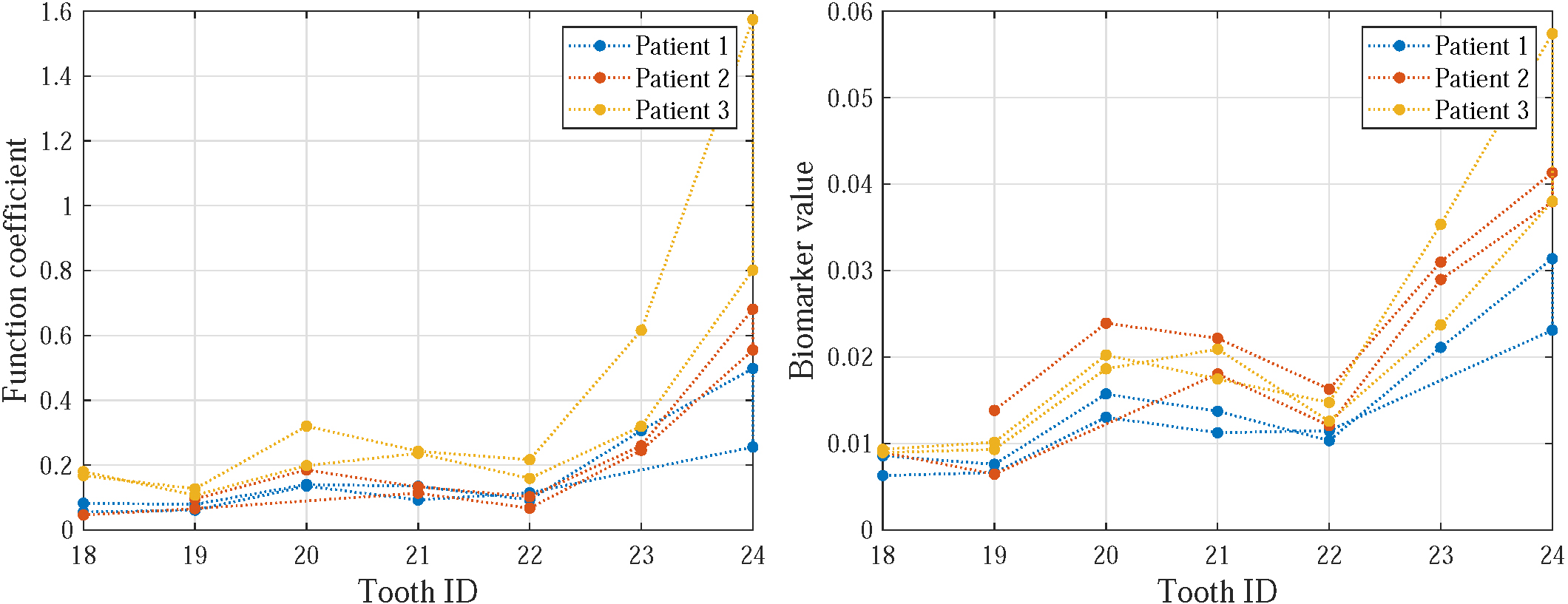}
 	\caption{Teeth movement variations of the three patients. \textbf{Left:} The coefficients of the functions fitted to the translation-load data. \textbf{Right:} The proposed biomarker values estimated for each patient's teeth. The right teeth IDs are reflected using the corresponding left teeth IDs, which results in two curves per patient.}
 	\label{fig:slopes}
 	\vspace{-0.25cm}
\end{figure}

To investigate the abovementioned assumption, first, we analyze the estimated coefficients of the fit functions ($\alpha_{t_{j,k}}$ and $\alpha_{\theta_{j,k}}$) for each tooth type. We observe that the teeth on the right side of the mandible show the same movement patterns as the corresponding teeth on the left side, where the UNN of the left side teeth can be calculated by subtracting the UNN of the corresponding teeth on the right side from 49. This provides us with more data points for the fitting purpose. The estimated coefficients of the nonlinear translation-load functions of the different patients are shown per tooth ID in \Cref{fig:slopes}. Note that the right teeth IDs are reflected in the same plot using the corresponding left teeth IDs.

\begin{figure}[t]
	\centering
	\includegraphics[width=\linewidth]{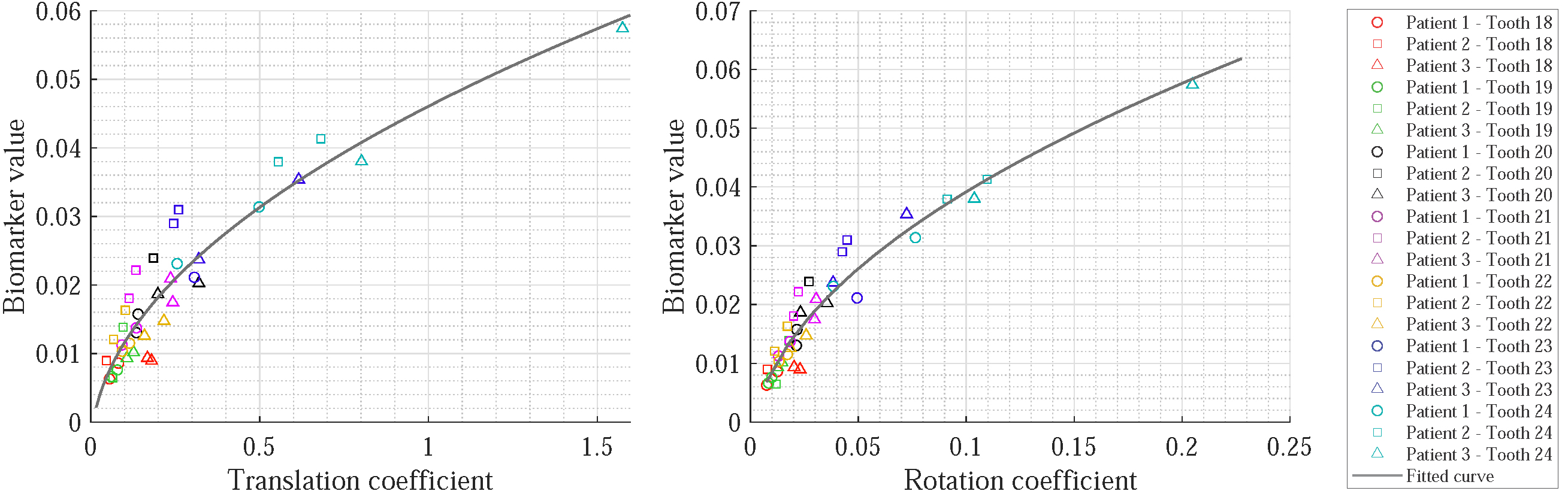}
 	\caption{The biomarker values versus coefficients of the displacement functions (translation and rotation). In each case, the behavior of the data is explained by a square root function.}
 	\label{fig:result_polys}
 	\vspace{-0.2cm}
\end{figure}

Second, the crown heights of teeth are extracted from the intraoral scan of each patient. These measured values are then divided by the root volumes of the corresponding teeth, in which the root volumes are calculated using the associated bounding boxes of the PDL geometries. \Cref{fig:slopes} illustrates the estimated biomarker values for each patient's tooth. The obtained ratios needs to be considered as a patient's tooth biomarker in the tooth displacement models of translation and rotation. Therefore, we investigate the relationship between the coefficients of the displacement functions and the proposed biomarker values. \Cref{fig:result_polys} shows the biomarker values versus coefficients of the teeth displacement functions (translation magnitude and rotation angle) for all patients' teeth. As it can be seen, the biomarker values and coefficients are in line with each other. For example, lower biomarker values and coefficients are associated with the molars while higher biomarker values and coefficients belongs to the incisors.

The relation between the biomarker values and coefficients of the teeth displacements can also be described by the square root functions, i.e., $b_t = \lambda_t \sqrt{\alpha_t} + \gamma_t$ and $b_{\theta} = \lambda_{\theta} \sqrt{\alpha_{\theta}} + \gamma_{\theta}$, where $b_t$ and $b_{\theta}$ are the biomarker functions associated with the translation/rotation function coefficients $\alpha_t$ and $\alpha_{\theta}$, respectively. Hence, the tooth displacements (translation/rotation) will be seen as a nonlinear function of both load ($l$) and the proposed biomarker ($b$), wherein the displacement-load function coefficients are replaced with the biomarker values. In other words, to obtain a patient's tooth displacements $t_{j,k}$ and $\theta_{j,k}$ for an applied load, one only needs to obtain the function coefficients $\alpha_{t_{j,k}}$ and $\alpha_{\theta_{j,k}}$ based on the biomarker value of the specific tooth using the fits shown in \Cref{fig:result_polys}.

\section{Summary and Conclusion}

The main goal of this work was to introduce a computational analysis tool for investigating the influence of the teeth geometry of different patients on the resulting teeth movements. Three biomechanical models were generated for studying the tooth movement variations of three patients. Our study showed that a combination of two clinical biomarkers, i.e., crown height and root volume could affect the tooth displacement. Therefore, we proposed two nonlinear functions for predicting translation and rotation of different patients' teeth for any applied load magnitudes. Proposing such functions not only allows for generalizability of the model across different patients but also provides a way to avoid having multiple values for different teeth IDs. To the best of our knowledge, this is the first time a full dentition intra-patient and inter-patient tooth movement analyses have been considered. This study focused on modeling the movement of teeth under an uncontrolled tipping scenario applied to three patients. The work still can benefit from investigating different tooth movement types such as the crown tipping, root tipping, and pure translation applied to some more patients.

\section{Acknowledgments}

\begin{minipage}{0.15\textwidth}
\vspace{-1.2cm}
\includegraphics[width=\textwidth]{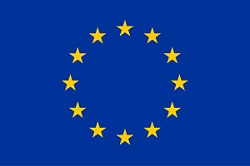}
\end{minipage}\hspace{10pt}
\begin{minipage}{0.8\textwidth}
This project has received funding from the European Union’s Horizon 2020 research and innovation programme under the Marie Sklodowska-Curie grant agreement No. 764644. This paper only contains the author's views and the Research Executive Agency and the Commission are not responsible for any use that may be made of the information it contains.
\end{minipage}

\FloatBarrier
\bibliographystyle{splncs04}
\bibliography{my_refs}

\end{document}